\documentclass[aps,prl,superscriptaddress,preprintnumbers,
showpacs,legalpaper,twoside,twocolumn]{revtex4}

\usepackage{bm}
\usepackage{graphicx}
\usepackage{amsfonts}
\usepackage{amsmath}
\usepackage{amssymb}

\begin{document}

\title{Probing correlated phases of bosons in optical lattices via trap
squeezing}

\author{Tommaso Roscilde}
\affiliation{Laboratoire de Physique, Ecole Normale Sup\'erieure de Lyon,
46 All\'ee d'Italie, 69007 Lyon, France}
\affiliation{Max-Planck-Institut f\"ur Quantenoptik, Hans-Kopfermann-strasse 1,
85748 Garching, Germany}

\begin{abstract}
We theoretically analyze the response properties of ultracold bosons 
in optical lattices to the static variation of the trapping potential.
We show that, upon an increase of such potential (trap squeezing), 
the density variations in a central region, with linear size of 
$\gtrsim 10$
wavelengths, reflect that of the \emph{bulk} system upon changing 
the chemical potential: hence measuring the density variations gives
direct access to the \emph{bulk compressibility}. When combined with 
standard time-of-flight measurements, this approach has the potential 
of unambiguously detecting the appearence of the most fundamental 
phases realized by bosons in optical lattices, with or without
further external potentials: superfluid, Mott insulator, band insulator
and Bose glass. 
\end{abstract}
\maketitle

 Ultracold gases in optical lattices offer the unique 
opportunity of literally implementing fundamental lattice 
models of strongly correlated quantum many-body systems, 
either bosonic or fermionic, traditionally considered as 
"toy" models for the description of complex condensed matter 
systems \cite{Fisheretal89, Scalapino95}. In the particular
case of ultracold bosons realizing the Bose-Hubbard (BH) model, 
recent experimental developments
have led to the spectacular demonstration of the Mott insulating (MI)
phase with controllable filling \cite{Greineretal02, Foellingetal06,
Campbelletal06}, and even more recent developments in laser
trapping offers the possibility of realizing further fundamental
insulating phases, such as a band insulator (BI) in a commensurate
superlattice \cite{commsup} or a Bose glass (BG) in an incommensurate
superlattice or in a laser-speckle potential \cite{speckles, Fallanietal07}. 

 Two main technical aspects limit to date the possibility of the 
experiments to retrieve full information on the true \emph{bulk} 
behavior of the model Hamiltonian implemented in the system. One aspect 
is the presence of a parabolic trapping potential which imposes
a spatial variation of the filling and hence a spatial modulation
of the local behavior exhibited by the system. A second aspect
is represented by the typical measurements performed on the 
system. 
The detection of strongly correlated phases is typically
based on the measurement of correlation functions (phase 
correlations via time-of-flight measurements \cite{Greineretal02} 
and density-correlations via noise-correlation analysis
\cite{Foellingetal05}) and on site-occupation statistics
\cite{Gerbieretal06, Campbelletal06}. Lattice-modulation spectroscopy
offers the possibility of measuring the dynamic structure
factor at zero transferred momentum \cite{Stoeferleetal04,Fallanietal07}
but it has the drawback of probing the \emph{global} response 
of the inhomogeneous system, and of being subject to a low-energy 
cutoff imposed by the duration of the experiment. This aspect 
prevents \emph{e.g.} the unambiguous observation of the BG, 
which does not have a special signature in correlation
functions, but it is unambiguously marked by the absence
of a gap in the excitation spectrum.

 The purpose of this paper is to propose a technique - 
\emph{trap-squeezing spectroscopy} - which
circumvents these two limitations at once, taking \emph{advantage}
of the parabolic trapping to extract the bulk behavior of the
model implemented in the system and in particular its lowest
particle-hole excitation energy. Two fundamental observations are 
at the basis of this proposal. On the one hand, the slowly
varying nature of the parabolic potential guarantees the 
validity of the local-density approximation (LDA)\cite{LDA, Roscilde08}, 
particularly
close to the potential minimum. Hence the density around the 
trap center mimics the behavior of the bulk system over an
\emph{extended} region of space of linear size of several 
($\gtrsim$10) lattice spacings, and consequently the average 
density in this region can be accessed via laser microscopy 
\cite{micro}. On the other hand, this central average density 
can be \emph{controlled} via the trapping potential
in very much the same way as the chemical potential 
controls the density of a bulk system in the grand-canonical
ensemble. In particular the trapping potential is by far
the lowest-energy potential to which the system is coupled
(with trapping frequencies as low as $\sim$10 Hz),
and measuring the response of the central density to small
variations of such potential allows to directly probe 
the low-energy response of the bulk system. 
 
 We theoretically investigate the trap-squeezing spectroscopy 
in the one-dimensional (1$d$) BH model in a parabolic potential 
\emph{plus} an external superlattice potential, 
${\cal H} (J,U,V_2,V_t) = {\cal H}_0(J,U,V_2) + 
V_t \sum_i (i-i_0)^2 n_i$ where
\begin{equation}
{\cal H}_0 = \sum_{i} \left[ -J
\left( b_i^\dagger b_{i+1} + \rm{h.c.} \right) +
\frac{U}{2} n_i(n_i-1) + V_2 ~g_i n_i
\right].
\end{equation} 
Here $ g_i = g_i(\alpha,\phi) = \cos^2(2\pi \alpha~i+\phi)-1/2$
is a one-color superlattice potential. 
$J$ and $U$ are experimentally controlled via the
height of the primary optical lattice, while $V_2$ is 
controlled via the height of a secondary optical lattice
\cite{Fallanietal07, commsup}. In the following we make
the fundamental assumption that $V_t$ can be 
varied \emph{independently} of the other parameters,
which is possible by applying an extra dipolar trap
to the system created by an additional running laser
wave.

 We study the above model via Stochastic Series Expansion quantum 
Monte Carlo \cite{SSE} at low temperatures 
(capturing the $T=0$ behavior) and in the grand-canonical 
ensemble, namely we simulate the Hamiltonian 
${\cal H}_{\mu} = {\cal H} - \mu \sum_i n_i$ where the chemical
potential $\mu$ is fine tuned to get the desired average
number of particles $\langle N \rangle$, and in this way
it becomes a function of the other Hamiltonian parameters
$\mu=\mu(V_t,N,J,U,V_2)$. According to LDA, the average density at
the center of the trap $n_C =: 1/|C|~\sum_{i\in C} \langle n_i \rangle$
(where the region $C$ will be defined later) reproduces closely 
that of a homogeneous system $(V_t=0)$ at a chemical potential
$\mu$. Hence controlling $\mu$ via one of the 
other parameters $V_t$, $N$, $J$, $U$, and $V_2$, allows to control
$n_C$. In particular, if $\mu$ is controlled by changing
$V_t$, namely by trap squeezing, while holding all the   
other parameters fixed, one has access to the 
\emph{compressibility} for the \emph{bulk} Hamiltonian
${\cal H}_0(J,U,V_2)$, estimated via $\kappa= \partial n_C/\partial \mu$. 

\begin{figure}[h]
\begin{center}
\includegraphics[
     width=68mm,angle=0]{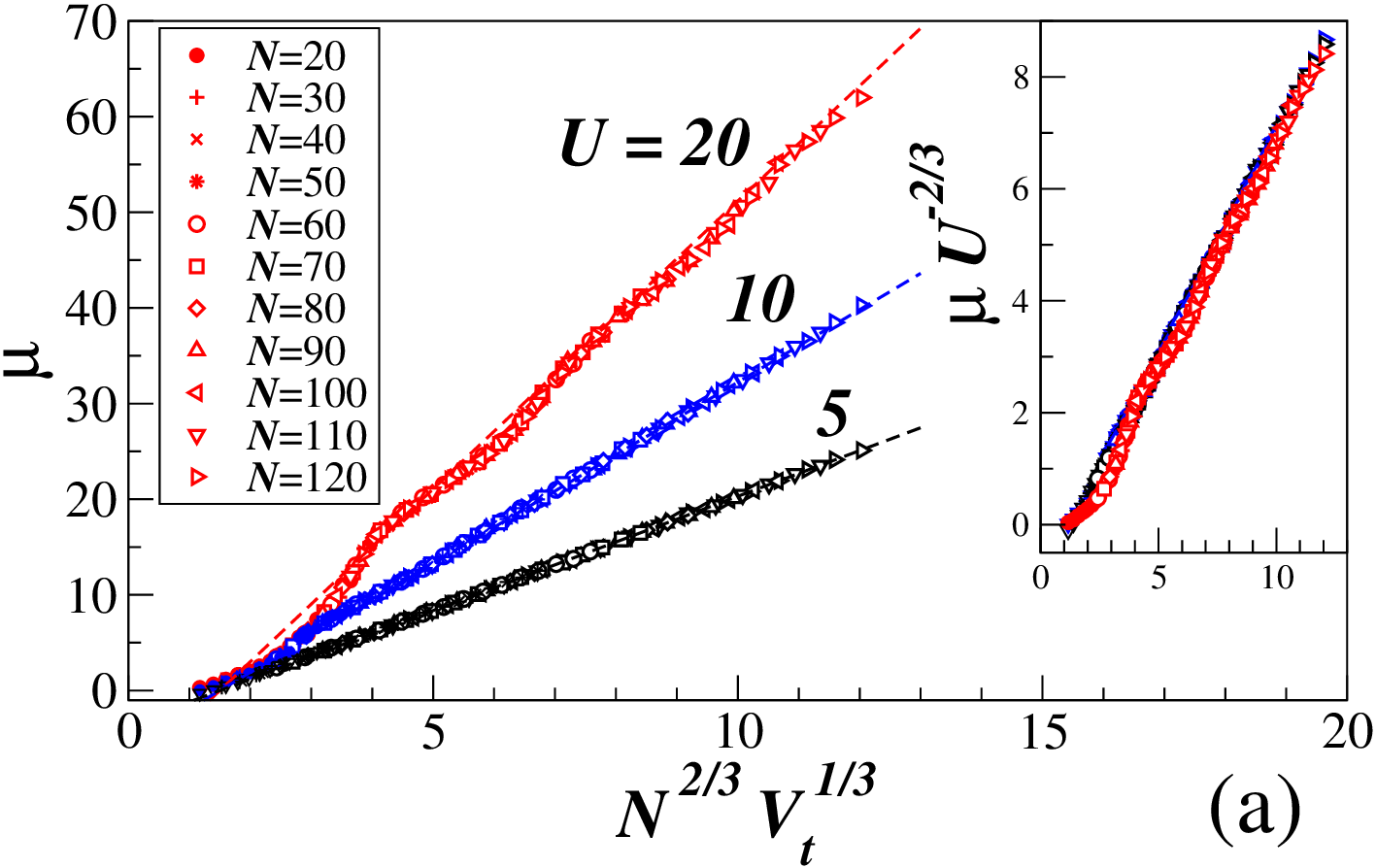} 
     \includegraphics[
     width=70mm,angle=0]{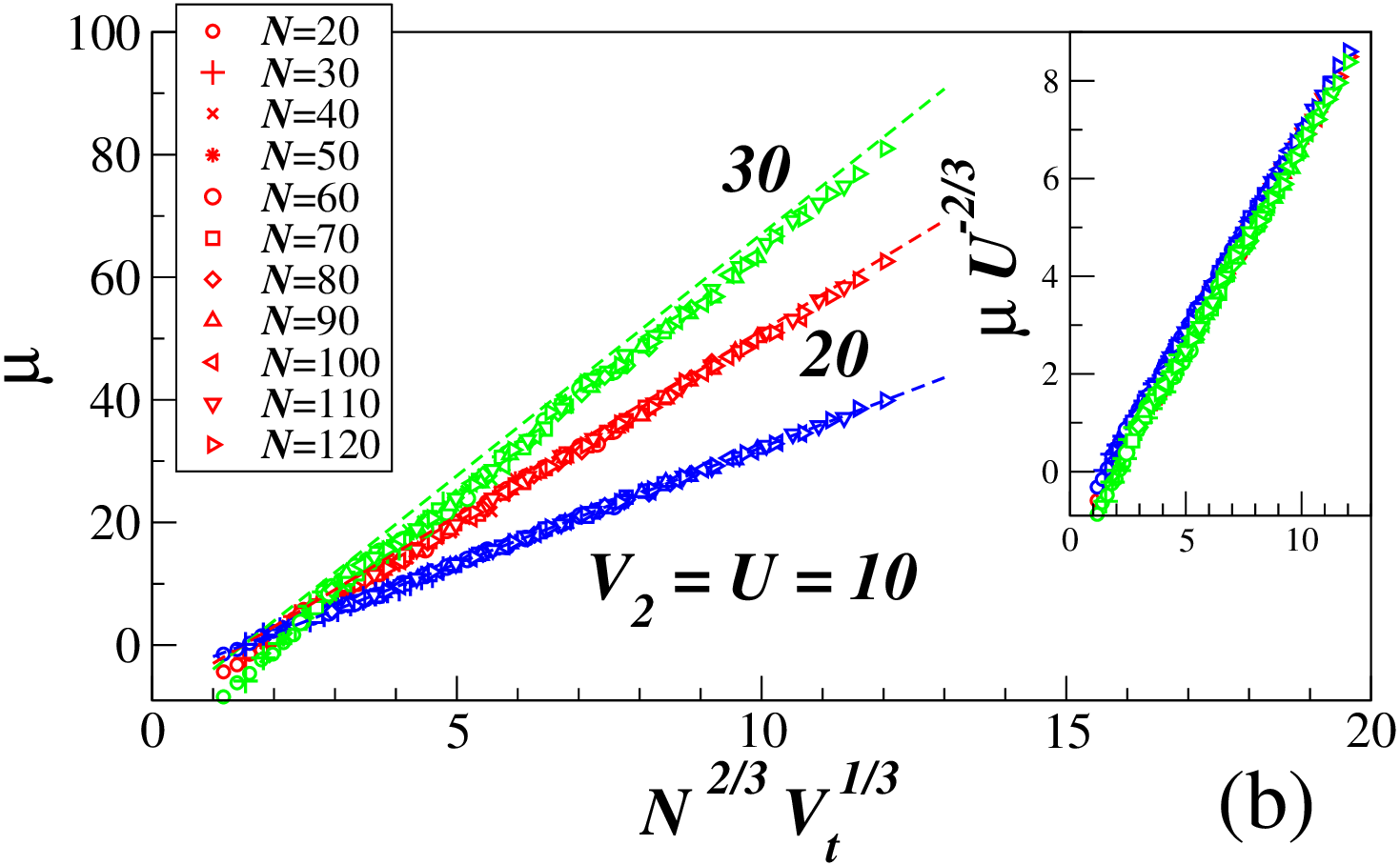} 
\caption{Scaling of the chemical potential $\mu$ for the 1d Bose-Hubbard 
model in a parabolic trap (a), and in a trap plus an incommensurate 
superlattice potential with strength $V_2=U$ (b). $\mu, V_t, V_2$ and $U$ 
are here reported in units
of $J$. The dashed lines correspond to a linear fit to 
the lowest-$(U/J)$ data, rescaled by a factor $(U/J)^{2/3}$ to 
compare with the other data sets. The inset shows the collapse
of all curves over a universal curve $f_\mu$ (see text).}
\label{f.mu}
\vspace*{-.5cm}
\end{center}
\end{figure} 

 The control on the chemical potential $\mu$ via trap squeezing 
 requires the detailed knowledge of the function 
 $\mu=\mu(V_t,N,J,U,V_2)$. Such a function can be accurately
 sampled via quantum Monte Carlo, given that its values
 are the result of the fine-tuning procedure of the chemical
 potential required to achieve a desired average $N$. 
 Fig.~\ref{f.mu} shows $\mu=\mu(V_t,N,...)$ for different 
 cases of the BH model without external potentials,
 $V_2=0$, and for an applied incommensurate superlattice potential 
 with strength $V_2=U$ and incommensurability parameter $\alpha=0.7714...$ 
 identical to that of the experiment of 
 Ref.~\onlinecite{Fallanietal07}. The behavior of $\mu$
 for a commensurate superlattice with $\alpha=3/4$ is
 found to be nearly identical to that of the incommensurate case. 
 In absence of a superlattice, 
 and for weakly interacting bosons, Thomas-Fermi (TF) theory 
 \cite{PethickS02} would predict the following scaling
 for the chemical potential in $d$-dimensions:
 $\mu \sim \left( N~ U\right)^{2/(2+d)} V_t^{d/(2+d)}$
 which gives $\mu \sim (N~U)^{2/3} ~V_t^{1/3}$ for $d=1$.
 Fig.~\ref{f.mu} shows that, at fixed $U$ and for \emph{all} the cases 
 considered, $\mu$ is a homogeneous function of the combination 
 $x=N^{2/3} ~V_t^{1/3}$; in particular, even for large $U/J$ 
 ratios it suprisingly verifies the TF prediction of a linear
 dependence on $x$; significant deviations are observed only 
 in the low-density and high-($U/J$) case, where the hardcore boson 
 regime sets in \cite{note.hardcore}. Moreover, for sufficiently low
 $U/J$ the data for different $U$'s collapse on the same 
 universal curve 
 \begin{equation}
 \mu = (U/J)^{2/3}f_{\mu}(x)
 \label{e.scaling}
 \end{equation}
 where $f_{\mu}$ is essentially a straight line. 
 A linear fit for the lowest-($U/J$) data gives 
 $f_{\mu}(x) = -1.225(13) + 0.817(2)x$ for \emph{all} the
 cases considered, namely in presence or in absence of
 a superlattice. We notice that the crude TF prediction
 would give $f_{\mu}(x) = [(d+2)\Gamma(d/2+1)/(2\pi^{d/2})]^{2/(2+d)}x$,
 which is well off the numerical data. 
 For higher $U/J$ we observe that the $(U/J)^{2/3}$
 scaling is still essentially obeyed by the derivative
 $\partial\mu/\partial x$ but not by the intercept
 $\mu(x=0)$; hence the slight disagreement between the scaling
 curve and the data for $U/J=20$ in Fig.~\ref{f.mu}(a) and
 $U/J=30$ in Fig.~\ref{f.mu}(b).
 
  Therefore we obtain a universal prediction for the dependence of 
 the effective chemical potential in the center of the trap
 on the experimentally controllable parameters $J$, $U$, $V_t$ and $N$
 for a large range of their values, and for the extreme case of 
 $d=1$ where the applicability of mean-field theory is in doubt. 
 Similar results
 are then expected to hold \emph{a fortiori} for the cases $d=2,3$.
 Hence we can firmly conclude that the chemical potential
 in the center of the trap represents a well controlled 
 \emph{experimental} parameter.
\begin{figure}[h]
\begin{center}
\includegraphics[
     width=65mm,angle=0]{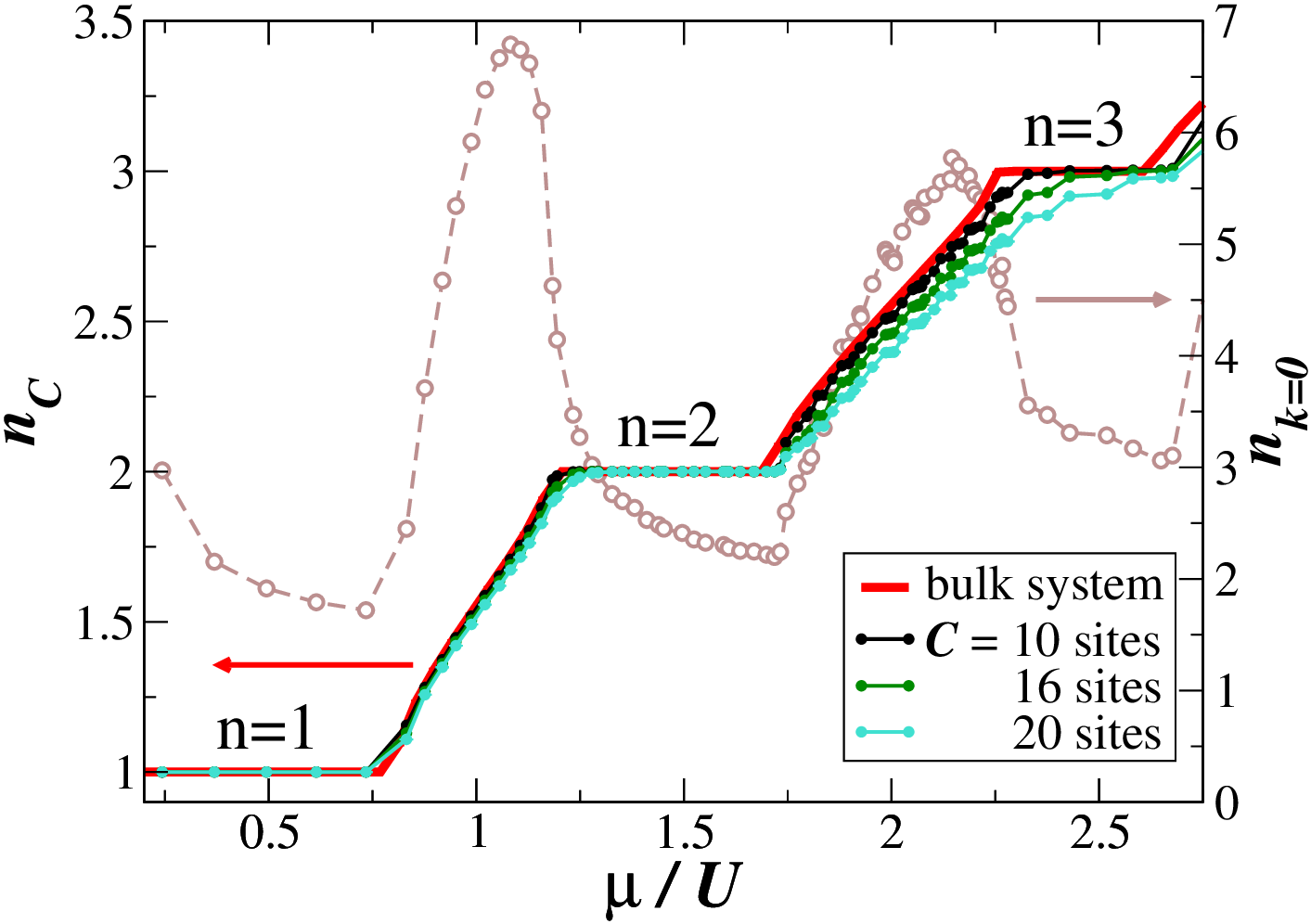}     
\caption{Central density $n_C$ and global coherent fraction
$n_{k=0}$ for the 1$d$ Bose-Hubbard model in a trap as a function 
of the effective chemical potential required to mantain $N=100$
bosons in the system. Here $U/J=20$. The 'bulk system' data
are obtained via a simulation on a homogeneous system 
with negligible finite-size effects and in the grand-canonical
ensemble with chemical potential $\mu$.}
\label{f.BH}
\vspace*{-.5cm}
\end{center}
\end{figure} 

  Armed with this prediction, we can then move on to simulate
 the outcome of a trap-squeezing experiment, where the 
 central density $n_C$ is monitored as a function of the trapping
 potential $V_t$. We start from the case of the 1$d$ BH model without
 any superlattice, for which we consider a boson number 
 $\langle N\rangle =100$ in a variable-frequency trap
 and with fixed repulsion $U/J=20$. 
 Fig.~\ref{f.BH} shows the evolution of the central density
 $n_C$ averaged over a region $C$ containing $10-20$ sites
 as a function of the chemical potential $\mu(V_t)$, and compared
 with the data for the bulk system. It is evident that, for
 a sufficiently low $\mu$ (namely for sufficiently
 low $V_t$), the bulk density curve is very well reproduced
 (in this case for $n_c\leq 2$). The deviation of $n_C$
 from the bulk value reveals that the truly homogeneous
 region in the trap center has become smaller than the 
 $C$ region, a fact that can be simply cured by increasing 
 the number of particles and decreasing the trapping
 potential so as to leave $\mu \sim V_t^{1/3} N^{2/3}$ 
 fixed. The succession of incompressible plateau regions
 at integer filling and compressible regions in the $n_c(\mu)$ 
 curve marks the alternation between incoherent MI and 
 coherent superfluid (SF) behavior, as also revealed by the (global) coherent fraction
 $n_{k=0} = (1/N) \sum_{ij} \langle b_i^{\dagger} b_j \rangle$.
 Remarkably, when the effective chemical potential 
 in the trap center overcomes the Mott gap, a few particles 
 can be transfered from the wings to the center into a locally
 SF state, and this gives rise to a violent increase in 
 the coherent fraction with a very sharp kink. 
 The width of the integer-filling
 plateaus corresponds to that of the MI lobes in 
 the phase diagram of the 1$d$ BH model: hence this kind of 
 measurement allows to reconstruct that phase diagram 
 with high accuracy, and to extract the 
 particle (hole) gap at any point as the minimal chemical
 potential variation required to increase (decrease) the density.  
 In particular, trap squeezing probes the density-driven 
 transition from MI to SF, which is in
 a different universality class \cite{Fisheretal89} with respect to 
 the transition driven by the $J/U$ ratio and probed so far
 in experiments \cite{Greineretal02, Stoeferleetal04,Spielmanetal07}. 
 Moreover we emphasize the high tolerance of the method to the 
 variation of the size of $C$, which corresponds to the size of 
 the focus of the imaging laser. 
 
\begin{figure}[h]
\begin{center}
\includegraphics[
     width=65mm,angle=0]{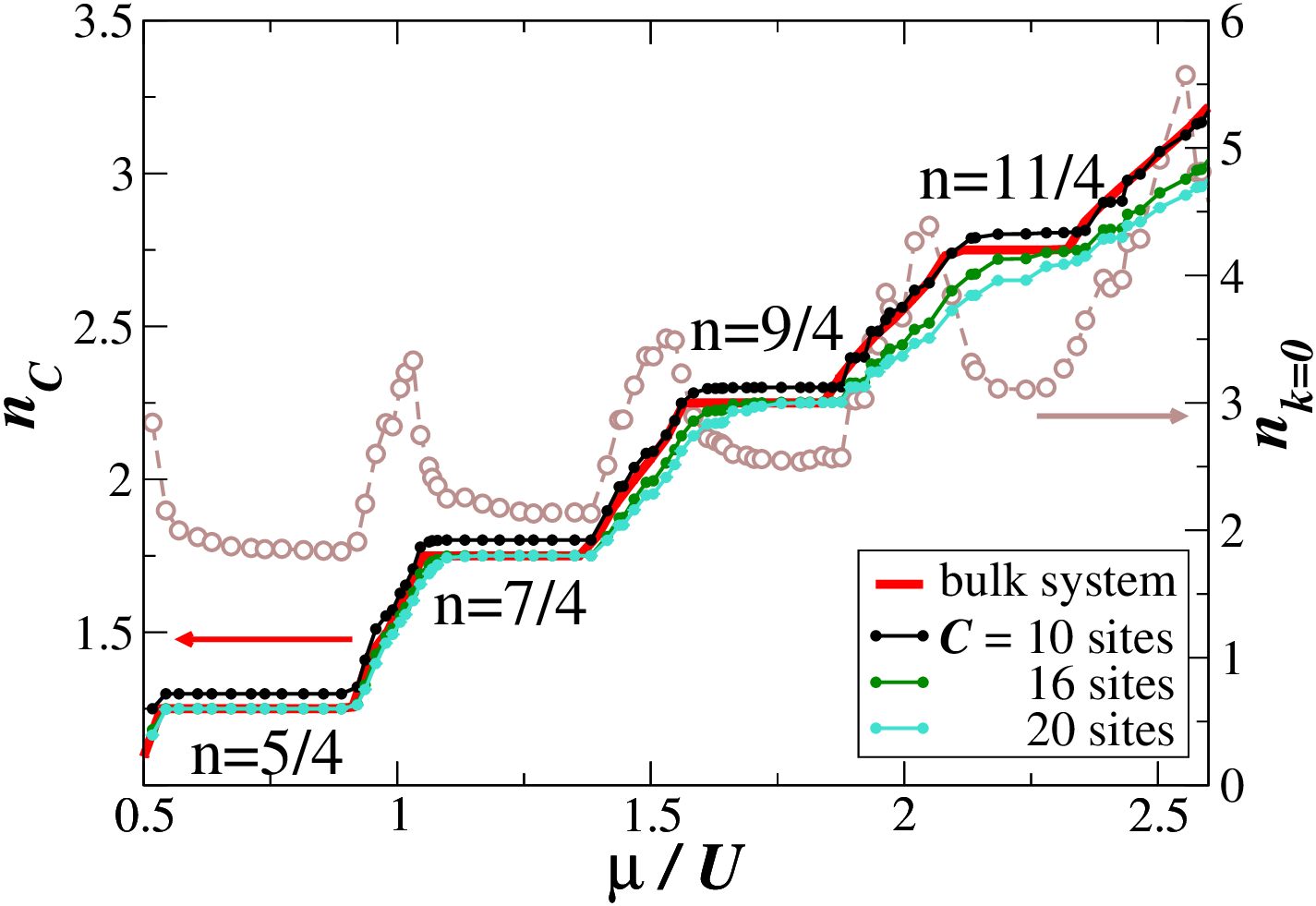}          
\caption{Central density and global coherent fraction
for the 1$d$ Bose-Hubbard model in a trap and in a commensurate
superlattice ($V_2=U=20 J$, $\alpha=3/4$, $\phi=0$). 
All symbols and notation as in Fig.~\ref{f.BH}.
Notice that the deviation of the data for $C=10$ sites from
the bulk ones is due to the fact that the $C$ region
does not contain an integer number of periods of the 
superlattice potential.}
\label{f.BI}
\vspace*{-.7cm}
\end{center}
\end{figure} 
 
 Having shown that trap squeezing allows to reconstruct
 the phase diagram of the \emph{bulk} Bose-Hubbard model, we 
 generalize this approach to probe other phases of correlated
 bosons in an optical lattice. To this end we consider $N=100$
 trapped bosons in an additional commensurate superlattice potential
 \cite{Rousseauetal06}
 with $\alpha=3/4$, fixed phase $\phi=0$, and strength $V_2=U=20J$, 
 such that it overcomes the MI gap and hence it removes the 
 MI phase; the insulating phase which is left for large
 $U/J$ is a BI with fractional, commensurate
 fillings $(2n+1)/4$ ($n=0,1,...$). Fig.~\ref{f.BI} shows
 the alternation of phases in the center of the trap 
 under trap squeezing as revealed by the central density,
 and compared to the bulk result; similarly to the MI-SF 
 transition, the BI-SF alternation is clearly evidenced. The 
 density plateaus correspond to the formation of incompressible 
 BI region in the trap center, an event associated with 
 a significant lowering of the global coherence in the system,
 as shown by the $n_{k=0}$ curve; the coherence is 
 suddenly increased when the BI gap is overcome by the chemical
 potential and particles are transfered into a locally SF
 state in the center.   
 
 \begin{figure}[h]
\begin{center}
\includegraphics[
     width=65mm,angle=0]{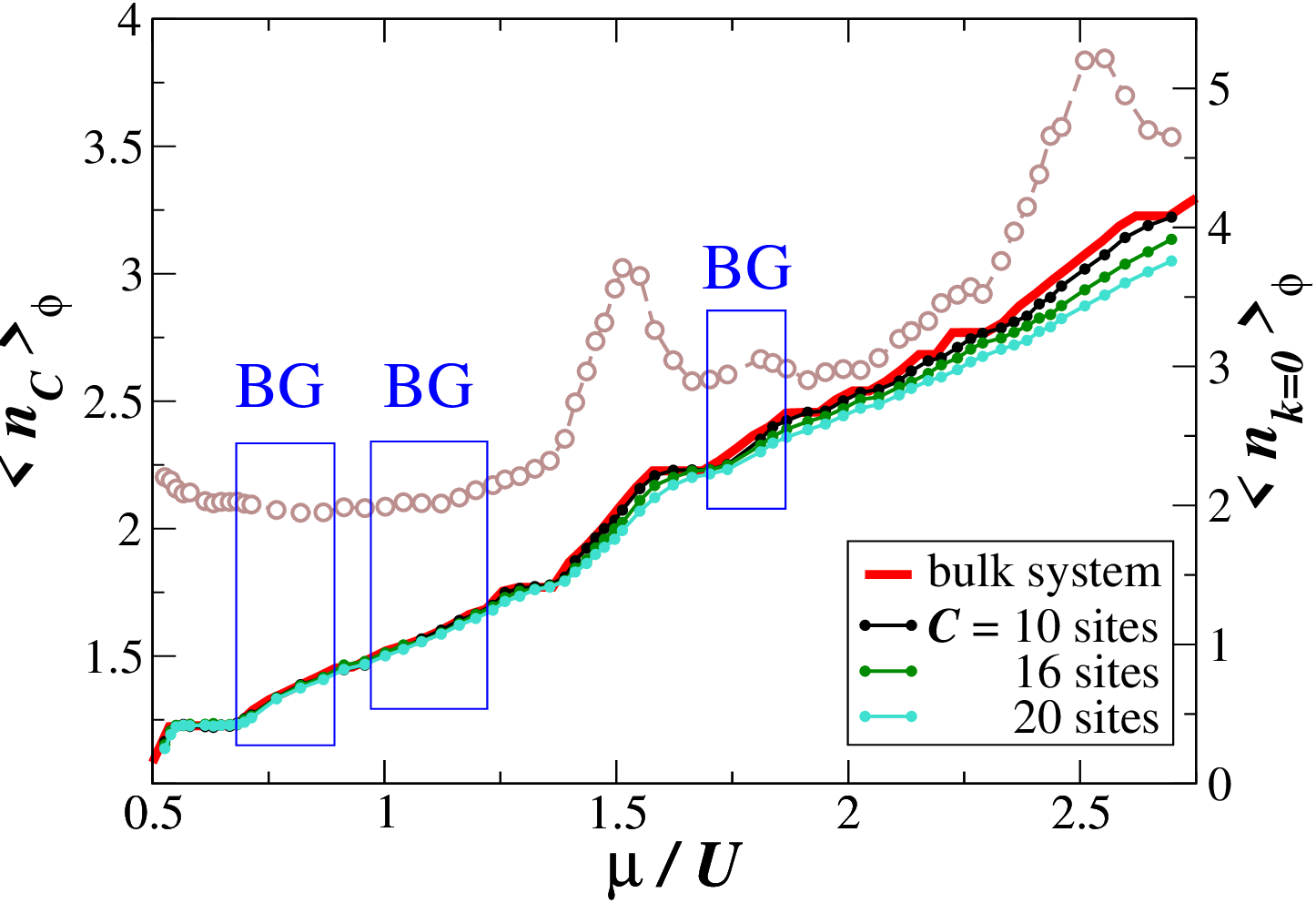}     
\caption{Central density and global coherent fraction
for the 1$d$ Bose-Hubbard model in a trap and in an incommensurate
superlattice ($V_2=U=20 J$, $\alpha=0.7714..$); $\langle...\rangle_{\phi}$
denotes the average over fluctuations of the spatial phase 
$\phi$. The boxes mark some relevant extended regions
exhibiting Bose glass (BG) behavior. 
All other symbols and notations as in Fig.~\ref{f.BH}.}
\label{f.BG}
\vspace*{-.5cm}
\end{center}
\end{figure}   
 
  The situation changes drastically when tuning slightly the
 superlattice parameter from the commensurate value $\alpha=3/4$ 
 to the incommensurate value $\alpha=0.7714..$ realized in recent
 experiments \cite{Fallanietal07}. In this case, 
 for a strong superlattice $V_2=U$ and for small $J/U$ the ground
 state of the system changes from SF to incompressible 
 incommensurate band insulator (IBI) and to compressible BG 
 upon changing the chemical potential \cite{Roscilde08}. 
 Fig.~\ref{f.BG} shows the variation under trap squeezing for
 the central density averaged over random fluctuations of the 
 spatial phase, $\langle n_C \rangle_{\phi}$. This average is
 intrinsic in current experimental setups, where the phase $\phi$
 can change from shot to shot, and it is essential for the 
 central region of the trap to sample 
 the full statistics of the quasi-periodic potential and hence 
 to mimic the bulk behavior of the system \cite{Roscilde08}. 
 Indeed we observe that $\langle n_C \rangle_{\phi}$  
 reproduces very well the bulk behavior for low enough density.
 In striking contrast to the previous two cases of no superlattice
 and of a commensurate superlattice, 
 the $\langle n_C \rangle_{\phi}$ curve exhibits extended compressible 
 regions for which the coherent fraction does \emph{not} vary upon changing 
 the chemical potential. This corresponds to transfer of particles 
 at no energy cost from the wings to the center of the trap 
 into \emph{localized} states which do not contribute to the 
 coherent fraction of the system: this fact provides smoking-gun 
 evidence for the appearence of a BG state in the center of the 
 trap \cite{estimates}. Moreover the joint information coming from the central
 density and the global coherent fraction enables to experimentally
 probe the incompressible IBI behavior and the compressible SF behavior. 

  In summary, we have proposed an experimental method  (trap squeezing 
  spectroscopy) to directly extract \emph{bulk} properties of 
  strongly correlated bosons from 
  measurements on a \emph{trapped} system - a fundamental 
  requirement in the future perspective of quantum 
  simulations of complex quantum systems realized with cold atoms.  
  The method relies
  on a simple, universal relationship between the trapping potential 
  and the effective chemical potential for the particles in 
  the trap center, which we numerically elucidate
  in the case of the Bose-Hubbard model realized in 
  optical lattices. 
  Measuring the response of the central density in the trap
  to the variation of the trapping potential provides direct
  access to the compressibility of the infinite system, a
  piece of information which is not directly accessible
  to current experimental setups and which is crucial to extract 
  the energy gap over the ground state of the Hamiltonian 
  implemented in the system. The method offers the possibility
  to extract the phase diagram of the Bose-Hubbard model
  with high resolution. Most remarkably, the joint measurement
  of the compressibility and of the coherent fraction (obtained
  via time-of-flight techniques \cite{Greineretal02}) provides
  clear evidence for the realization of a Bose-glass
  state in the center of the trap. We have demonstrated this 
  property in the case of an incommensurate superlattice as recently
  realized in experiments \cite{Fallanietal07}, although the
  same technique can be applied to different realizations of
  random or pseudo-random potentials \cite{speckles}.
  
  From the experimental point of view, this method requires the
  application of an extra dipolar trap whose strength can be
  controlled independently of that of the optical lattice, 
  and the measurement of the optical depth of the cloud
  over a region of order $\sim 10$ wavelengths ($\sim ~10 \mu$m) 
  of the optical lattice in all three spatial directions. The 
  measurements of the central density and of the coherent fraction 
  cannot typically be performed in the same shot, so that special
  care is needed in mantaining the number of particles $N$
  fixed from shot to shot to achieve the same experimental
  conditions. This can be typically achieved by post-selecting
  only those measurements with the same total $N$ in the trap.  
  Useful discussions with N. Bar-Gill, L. Fallani, C. Fort, and M. Rigol
  are gratefully acknowledged.


\begin{thebibliography}{99}
\bibitem{note.hardcore} In this regime of low density and 
high $U/J$ the behavior of the system
becomes independent of $U$ as observed in Fig.~\ref{f.mu}(a), given
that multiple occupancy is essentially
absent in the system. In presence of a strong superlattice $V_2=U$, 
on the contrary, multiple occupancy persists at lower densities due
to confinement of the particles in the superlattice minima.
This explains the absence of the hardcore regime in the
data of Fig.~\ref{f.mu}(b).
\bibitem{estimates} An estimate of the experimental parameters 
\cite{Roscilde08}
corresponding to the regions marked as BG in Fig.~\ref{f.BG}
at the lowest chemical potentials gives a trapping frequency
$\nu\approx 30-50$ Hz for $^{87}$Rb in a one-dimensional
optical superlattice: here we consider transverse standing waves of 
wavelength  $\lambda_1 = 830$ nm and intensity 
$V_{\perp}=30 E_r$, a longitudinal primary one (at $\lambda_1$) of 
$V_{0}\approx 8.5 E_r$ and a secondary one (at $\lambda_2=1076$ nm as in 
Ref.~\onlinecite{Fallanietal07}) with $V_{0}'\approx 0.6 E_r$
giving $V_2 \approx U\approx 20 J$ 
($E_r=h^2/2m\lambda_1^2$ is the recoil energy).
Hence, the the observation of BG behavior is possible
for a fully realistic set of parameters.
\end{thebibliography}
\end{document}